\documentclass[preprint,amsmath,amssymb,floatfix]{revtex4}

\usepackage{times}
\usepackage{graphicx}
\usepackage{dcolumn}
\usepackage{bm}
\usepackage{psfrag}

\begin{document}

\preprint{APS/123-QED}

\title{Structure and stability of helices in square-well homopolymers}

\author{M.~N.~Bannerman}
\author{J.~E.~Magee}
\author{L.~Lue}
\email{leo.lue@manchester.ac.uk}
\affiliation{
  School of Chemical Engineering and Analytical Science \\
  The University of Manchester \\
  PO Box 88, Sackville Street \\
  Manchester M60 1QD \\
  United Kingdom
}

\date{\today}

\begin{abstract}
Recently, it has been demonstrated [Magee et al., {\em
  Phys.\ Rev.\ Lett.\/} {\bf 96}, 207802 (2006)] that isolated,
square-well homopolymers can spontaneously break chiral symmetry and
``freeze'' into helical structures at sufficiently low temperatures.
This behavior is interesting because the square-well homopolymer
is itself achiral.  In this work, we use event-driven molecular
dynamics, combined with an optimized parallel tempering scheme, to
study this polymer model over a wide range of parameters.
We examine the conditions where the helix structure is
stable and determine how the interaction parameters of the polymer
govern the details of the helix structure.
The width of the square well (proportional to $\lambda$) is found to
control the radius of the helix, which decreases with increasing well
width until the polymer forms a coiled sphere for sufficiently large
wells.
The helices are found to be stable for only a ``window'' of molecular
weights.  If the polymer is too short, the helix will not form.  If
the polymer is too long, the helix is no longer the minimum energy
structure, and other folded structures will form.  The size of this
window is governed by the chain stiffness, which in this model is a
function of the ratio of the monomer size to the bond length.
Outside this window, the polymer still freezes into a locked structure
at low temperature, however, unless the chain is sufficiently stiff,
this structure will not be unique and is similar to a glassy state.
\end{abstract}

\pacs{Valid PACS appear here}

% \keywords{Suggested keywords}

\maketitle

%%%%%%%%%%%%%%%%%%%%%%%%%%%%%%%%%%%%%%%%%%%%%%%%%%%%%%%%%%%%%%%%%%%%%%%%%%%%%%%%
%%%%%%%%%%%%%%%%%%%%%%%%%%%%%%%%%%%%%%%%%%%%%%%%%%%%%%%%%%%%%%%%%%%%%%%%%%%%%%%%
\section{Introduction}

One of the fascinating features of proteins is their ability to lock
into a specific, folded structure.  This feature is often crucial to
their function.  A key structural unit which frequently appears in
proteins is the helix.  Helical structures also appear in other
molecules, such as in DNA, homopolypeptides (e.g., polyalanine), as
well as in some synthetic polymers.
Consequently, there has been a lot of interest in the helix-coil
transition as a starting point to understanding the more general issue
of protein folding.

Many detailed computer simulations on ``realistic'' interaction
potential models have been conducted, to better understand the
formation of helices in polypeptides and proteins (e.g., see
Refs.~\cite{Okamoto_Hansmann_1995,Brooks_2002,Nguyen_etal_2004,Scheraga_etal_2007}).
In these systems, the formation of hydrogen bond interactions between
different amino acid groups is principally responsible for the
formation of the helix.
Helices also spontaneously form in simplified interaction models that
have short-ranged, directional interactions between their constituent
monomers \cite{Kemp_Chen_1998,Kemp_Chen_2001}.
Many theories have been developed to describe the helix-coil
transition in homopolypeptides and other biological molecules,
starting with the pioneering work of Zimm and Bragg
\cite{Zimm_Bragg_1959} and later followed by many others
\cite{Schellman_1958,Lifson_Roig_1961,Munoz_Serrano_1994,Doig_2002,Varshney_etal_2004}.
The key feature of these theories is the characterization of a
distinct helix and coil state for each residue in the peptide chain.
This is justified for these systems because of the specific
arrangement of the residues in the helix conformation and the large
energies due to the formation of the hydrogen bonds.
While these approaches have led to keen insights for helix formation
in polypeptide and protein molecules, they are dependent on the fact
that short-ranged, directional interactions drive the formation of the
helix structure.  In these molecules, one can argue that the helix
structure has been ``built'' in.

Can the helix structure occur in molecules without these specific
interactions, and if so, what then controls its geometry?  It has been
suggested that the helix is a stationary configuration for
semi-flexible chains~\cite{Kholodenko_etal_1998}, and the optimal
shape of flexible~\cite{vogel_etal_2009} and closely packed,
compact~\cite{Maritan_etal_2000,Marenduzzo_etal_2005} strings.  This
hints at a more general driving force for helix formation in real
proteins and may explain why the structure is so prolific in nature.
In order to gain some more general understanding of the mechanisms
behind helix formation, we examine the square-well homopolymer model.
This is a simple polymer model composed of linearly bonded, hard
spheres that interact with each other through an isotropic square-well
attraction.  Isolated square-well homopolymers exhibit the typical
coil to globule transition observed in many polymers as the
temperature is decreased below the theta point; however, they
also freeze into compact, crystal-like structures
\cite{Zhou_etal_1996,Zhou_etal_1997} at sufficiently low temperatures.
Interestingly, Magee et al.\ \cite{Magee_etal_2006} have demonstrated
that, by introducing stiffness, the square-well homopolymer model can
fold into a helix structure.  This is a remarkable result, as the
model is achiral and yet it spontaneously breaks symmetry and folds
into left or right handed helices.  This is merely a result of the
polymer being stiff, having an excluded volume and an attractive self
interaction.
An exact analysis of the density of states of square-well tetramers
and pentamers was performed \cite{Magee_etal_2008} to examine the
relationship between the distributions and correlations of the
torsional angles in these fragments to the stability of the helix in
longer length chains.  However, the question still remains as to what
controls the geometry and the stability of the helical structures
formed by these molecules.

In this work, we use molecular dynamics, combined with the replica
exchange method, to explore the behavior of square-well homopolymers
to better understand the link between the interactions between the
monomers of the chain and the overall structure of the molecule.  In
particular, we are interested in the range of conditions over which
the helix structure is stable.
The remainder of this paper is organized as follows.
Section~\ref{sec:sim} describes the details of the square-well
homopolymer model that we investigate in this work.  In addition, it
provides background information on the simulation methods we employed
and outlines the procedures used to generate and analyze the resulting
simulation data.
The results of the simulations are presented in
Section~\ref{sec:parameters}.  This section begins with an overview of
the general behavior exhibited by the square-well homopolymers.  Then,
it continues by analyzing and discussing the influence of the bond
length (or equivalently monomer size), the range of attraction between
monomers, and the total number of monomers in the polymer on the
structure and thermodynamic behavior of the homopolymer.
Finally, the major findings of this work are summarized in
Section~\ref{sec:conclusions}.

%%%%%%%%%%%%%%%%%%%%%%%%%%%%%%%%%%%%%%%%%%%%%%%%%%%%%%%%%%%%%%%%%%%%%%%%%%%%%%%%
%%%%%%%%%%%%%%%%%%%%%%%%%%%%%%%%%%%%%%%%%%%%%%%%%%%%%%%%%%%%%%%%%%%%%%%%%%%%%%%%
\section{\label{sec:sim} 
Simulation details}

\begin{figure}[tb]
\centering
\includegraphics[clip,width=\columnwidth]{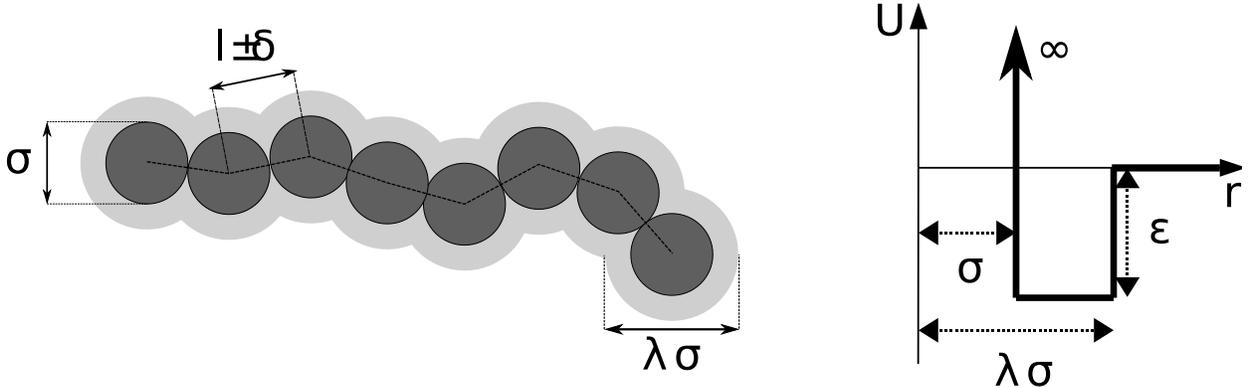}
\caption{\label{fig:helixmodel}Polymer model with bond length
  $l\pm\delta$, well energy $\epsilon$, well width $\lambda\sigma$,
  and monomer diameter $\sigma$.  The interaction energy between two
  non-bonded monomers separated by a distance $r$ is given on the
  right.}
\end{figure}

The polymer model that we study in this work is a chain of linearly
bonded monomers.  
Monomers that are not directly bonded together interact with each
other through the potential
\begin{equation}
u(r) = \left\{
\begin{array}{rl}
   \infty & \quad\mbox{for~~$r < \sigma$} \\
-\epsilon & \quad\mbox{for~~$\sigma < r < \lambda\sigma$} \\
        0 & \quad\mbox{for~~$\lambda\sigma < r$} \\
\end{array}
\right.
\end{equation}
where $r$ is the distance between the centers of the monomers.  Each
monomer is a hard sphere of diameter $\sigma$, surrounded by an
attractive square well of diameter $\lambda\sigma$.  When two monomers
are within a distance $\lambda\sigma$, they feel an attractive
interaction energy of magnitude $\epsilon$.
Monomers that are directly bonded together interact with each other
through the potential
\begin{equation}
u_{\rm bond}(r) = \left\{
\begin{array}{rl}
\infty & \quad\mbox{for~~$r < l-\delta           $} \\
0      & \quad\mbox{for~~$l-\delta < r < l+\delta$} \\
\infty & \quad\mbox{for~~$l+\delta < r           $} \\
\end{array}
\right.
\end{equation}
The bond length is nominally equal to $l$ but is allowed to fluctuate
between $l-\delta$ and $l+\delta$.
If $\sigma/l>1$, then directly bonded monomers in the chain overlap.
Monomers that are not directly bonded together other are not allowed
to overlap each other.  This induces a stiffness in the polymer, due
to the restrictions on the allowed bond angles imposed by the excluded
volume interaction between monomers separated by two bonds.  In the
limit that $\sigma/l$ approaches $2(1+\delta/l)$, the chain becomes
completely rigid.  For all the simulations presented here,
$\delta/\sigma=0.1$, and so the bond length is allowed to vary by
$\pm10\%$.
A schematic drawing of the polymer model is given in
Fig.~\ref{fig:helixmodel}.

%%%
%%%
%%%
%%%

We use constant temperature molecular dynamics (MD) to investigate the
structural and thermodynamic properties of the square-well polymer
chains over a range of temperatures.
The temperature of the simulations was maintained with the Andersen
thermostat \cite{Andersen_1980}.
The basic algorithm that we employ to perform the MD simulations is
based on the one originally developed by Alder and Wainwright
\cite{Alder_Wainwright_1959}.
Several subsequent advances have significantly improved the
computational speed of this original algorithm.  These include the use
of overlapping cells \cite{Rapaport_1980,Krantz_1996}, the delayed
states algorithm \cite{Marin_etal_1993}, and calender event queues
\cite{Paul_2007}.
We have incorporated these advances in order to construct a code where
the computational cost of the simulation is independent of the number
of particles $N$ in the system.

One shortcoming of molecular dynamics is that it is prone to becoming
trapped in local energy minima, especially at low temperatures.  In
particular for conditions where helical or other ``frozen'' structures
are formed, the homopolymer may become locked within a specific
configuration.  Using only molecular dynamics, the helices formed by
the square-well polymers are stable over the length of accessible
simulation times and rarely transform between the left and right
handed forms.  This makes the study of the equilibrium behavior of
these systems at low temperatures extremely formidable.

To overcome this difficulty, the MD simulations are coupled with the
replica exchange/parallel tempering method \cite{Swendsen_Wang_1986}.
In this technique, several molecular dynamics simulations, each at a
different temperature, are run simultaneously; a Monte Carlo move is
added to exchange chain configurations between simulations at
different temperatures.  A configuration that is locked at a low
temperature may then move up in temperature, unfold, and drop in
temperature to sample another configuration.  This enables the systems
to rapidly overcome local energy minima and better explore the full
range of available configurations.

\begin{figure}[tb]
\begin{center}
\includegraphics[clip,width=\columnwidth]{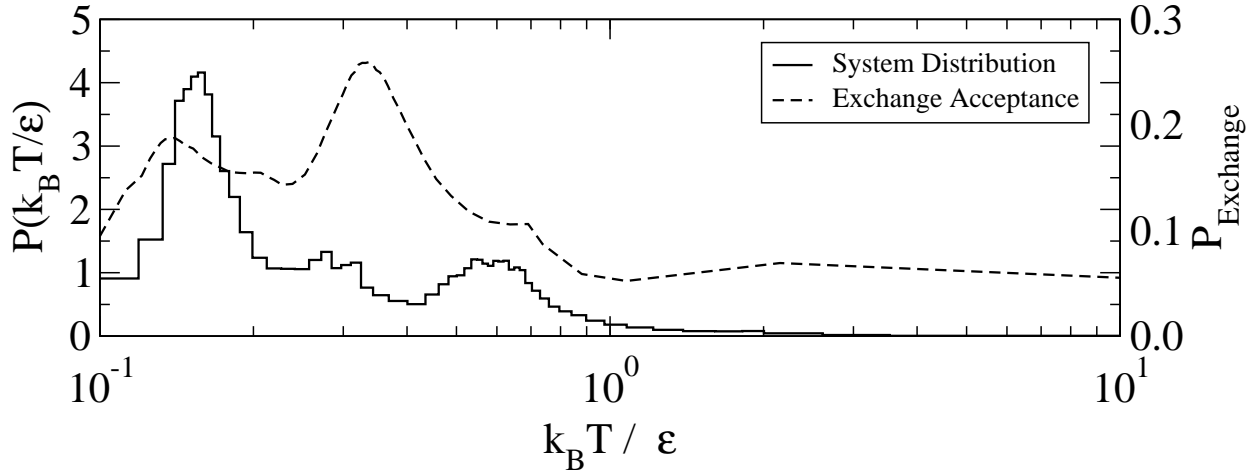}
\caption{\label{fig:replicaExInfo}
Replica exchange simulations for an isolated square-well homopolymer
with $N=20$, $\sigma/l=1.6$, and $\lambda=1.5$.  Optimal distribution
of system temperatures is given by the solid line, and the acceptance
ratio of the replica exchange move for adjacent temperature systems is
given by the dashed line. }
\end{center}
\end{figure}

The effectiveness of the replica exchange method depends on the choice
of the temperatures of the individual simulations.  In order to
determine the optimal values of these temperatures, we use an approach
recently developed by Katzgraber et al.\ \cite{Katzgraber_etal_2006}.
This maximizes the number of configurations that travel between the
lowest and highest temperature simulations, as modeled by a
one-dimensional diffusion process.  A typical optimal distribution of
system temperatures is presented in Fig.~\ref{fig:replicaExInfo},
along with the resulting exchange rates.  The optimization procedure
clusters the simulation temperatures near conditions where the polymer
undergoes structural changes with significant topological differences.
The optimal distribution of system temperatures does not correspond to
a constant acceptance ratio \cite{Katzgraber_etal_2006}, as is
commonly presumed.

%%% 
%%% 
%%%
%%% 

A series of $NVT$ molecular dynamics simulations, combined with the
replica exchange method, is performed to examine the properties of
square-well homopolymers over a range of values for $\sigma/l$,
$\lambda$, and $N$.  
For each particular chain, $51$ temperatures are used, and the systems
are equilibrated for $10^4$ attempted replica exchange moves.  The
replica exchange move consists of selecting $5\times51$ random pairs
and attempting to swap the configuration between each pair.  Between
each replica exchange move, the dynamics of the isolated polymers is
run for a few hundred mean free times.  Following an initial
equilibration period, data are collected over $5\times10^4$ attempted
replica exchange moves. The collected data are then interpolated using
multiple histogram reweighting~\cite{Ferrenberg_Swendsen_1989} to
obtain smooth heat capacity curves as a function of the temperature.

%%%%%%%%%%%%%%%%%%%%%%%%%%%%%%%%%%%%%%%%%%%%%%%%%%%%%%%%%%%%%%%%%%%%%%%%%%%%%%%%
%%%%%%%%%%%%%%%%%%%%%%%%%%%%%%%%%%%%%%%%%%%%%%%%%%%%%%%%%%%%%%%%%%%%%%%%%%%%%%%%
\section{Results and discussion
\label{sec:parameters}}

%%%%%%%%%%%%%%%%%%%%%%%%%%%%%%%%%%%%%%%%%%%%%%%%%%%%%%%%%%%%%%%%%%%%%%%%%%%%%%%%
\subsection{Overview}

\begin{figure}[tb]
\centering
\includegraphics[width=\columnwidth,clip]{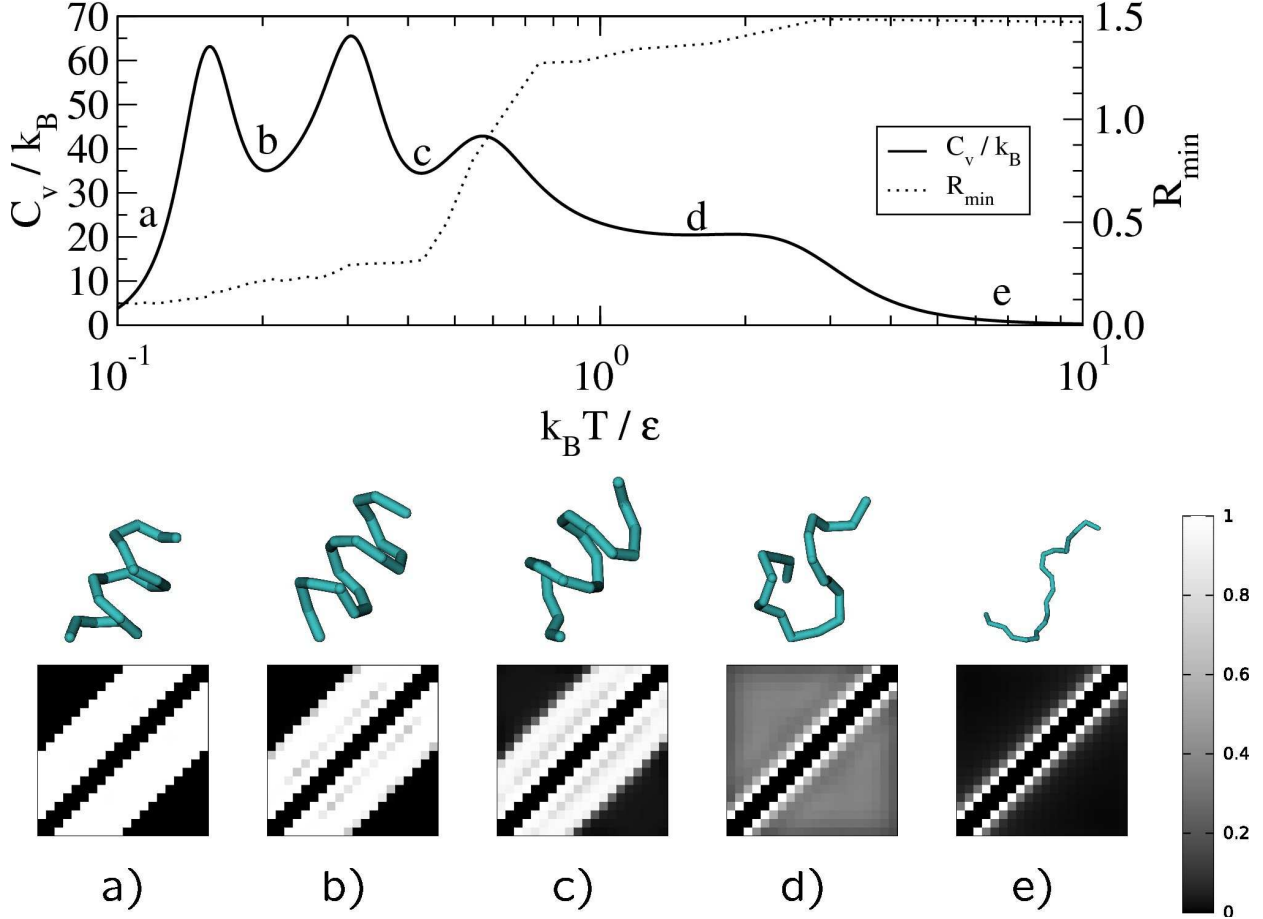}\\
\caption{\label{fig:initResults}
The heat capacity, the optimal configuration RMSD,
sample configurations, and contact maps at various temperatures
($a$--$e$) of an isolated helix homopolymer with $N=20$,
$\sigma/l=1.6$, and $\lambda=1.5$.}
\end{figure}

To illustrate the general behavior of the square-well homopolymers, we
present results from MD simulations in Fig.~\ref{fig:initResults}
for a chain consisting of $N=20$ monomers with $\sigma/l=1.6$ and
$\lambda=1.5$.  The solid line in the plot shows the variation of the
excess heat capacity $C_v$ with temperature.  The peaks of the heat
capacity typically indicate transitions between different structural
states of the polymer.

To characterize the rigidity (i.e., ``frozen'' vs.\ flexible) of the
structure of the homopolymer at a particular temperature, we collect
$N_{ss}$ configurations of the polymer at regular intervals in time
during the course of the simulation (here we choose 10 replica
exchange times).  For each sampled configuration $\alpha$, we then
determine the average $R_\alpha$ of the root mean square difference
(RMSD) against all other collected configurations, which is given by
\begin{align}
\label{eq:Ralpha}
R_{\alpha} = N_{ss}^{-1} \sum_{\alpha'=1}^{N_{ss}} {\rm RMSD}(\alpha,\alpha'),
\end{align}%
where the RMSD between two configurations $\alpha$ and $\alpha'$ is
defined as
\begin{align}
\label{eq:rmsd}
{\rm RMSD}(\alpha,\alpha') = \left[ N^{-1} \sum_{i=1}^{N}
  \left| {\bf  r}_{i}^{(\alpha)} - {\bf r}_{i}^{(\alpha')} \right|^2
\right]^{1/2}.
\end{align}%
and ${\bf r}_{i}^{(\alpha)}$ is the position of monomer $i$ in the
polymer of configuration $\alpha$. The reported value of the RMSD
between a pair of configurations is the minimum value obtained by
rotating~\cite{Kabsch_1976} and reflecting the configurations, as well
as reversing the numbering sequence of the monomers.  We consider the
configuration with the lowest value of $R_\alpha$ as the most
representative of the entire set of sampled configurations of the
homopolymer.
%The configuration with the lowest value of $R_\alpha$ is the closest
%to describing the average structure of the set.
The average RMSD of this configuration, denoted by $R_{\rm min}$, is
used to indicate how rigid the polymer structure is at a given
temperature (i.e., $R_{\rm min} = \min_{\alpha} R_{\alpha}$).
Low values of $R_{\rm min}$ suggest that the homopolymer remains
``frozen'' within same structural configuration.  
High values of $R_{\rm min}$ indicate that the homopolymer is not well
characterized by a single structure.  This can imply that the
homopolymer is in a rather flexible state, such as a coil or a molten
globule.  However, high values of $R_{\rm min}$ could also result if
the homopolymer can be ``frozen'' into several distinct
configurations, such as in a glassy state.  Using cluster analysis of
the distance matrix formed by the RMSD's of every pair of sample
configurations, it is possible to estimate the number of stable states
and thereby distinguish between these two situations.  This issue will
be discussed further in Sec.~\ref{sec:length}, where the effect of
polymer length is explored.

%%%
%%%
%%%
%%%

The variation of $R_{\min}$ with temperature is given by the dotted
line in the plot in Fig.~\ref{fig:initResults}.
Beneath the plot are sample configurations of the homopolymer at
several different representative temperatures.  Underneath each of
these configurations is the corresponding contact map of the average
structure, which details the proximity of pairs of monomers in the
polymer.  The positions along the ordinate and abscissa of the contact
maps denote each of the monomers along the chain.  The locations
within the contact map are shaded according to how often two monomers
interact with each other (i.e.\ within a distance $\lambda\sigma$).
Black denotes no interaction, white denotes continuous interaction,
and gray denotes intermittent interaction.  For monomers that are
bonded together (along the diagonal), we have shaded the entries in
the contact map black.  For the case where $\sigma/l=1.6$ and
$\lambda=1.5$, monomers that are separated by two bonds are always in
each others attractive well due to the overlap and result in two
off-diagonal white bands.

At higher temperatures (point $e$), the polymer is extended, and the
contact map indicates that the monomers of the polymer are rarely in
contact with each other.  The optimal configuration RMSD ($R_{\rm min}$)
also indicates that the typical configuration is not locked but,
instead, is quite flexible.  Upon decreasing the temperature, a
shoulder in the heat capacity marks the transition from an extended
coil to a globule state $d$.  The contact map indicates that while
monomers do interact significantly with each other, they do not remain
in continuous contact with the same monomers, and, therefore, the
contact map is primarily gray.  Although the polymer has collapsed
into a compact structure, it contains no regular structure and
$R_{\rm min}$ remains high.

Decreasing the temperature still further (point $c$), we see that the
polymer changes from an unstructured globule to a more ordered helical
structure.  The contact map shows the stripe pattern that is
characteristic of a spiral or helical structure.  Two more helical
structures are present at lower temperatures (points $b$ and $a$)
which possess a slightly different pitch and radius.  The value of
$R_{\rm min}$ decreases sharply over the first helix transition as the
polymer forms a regular structure.  This decreases further, indicating
that the structures become more rigid.  This is in agreement with the
contact maps, where intermittent contacts become permanently ``on'' at
low temperatures.
Points $c$ and $b$ correspond to the helix 1 and 2 structures in the
diagram of states presented by Magee et al.\ \cite{Magee_etal_2006}.
We will refer to the structure at point $a$ as the helix 3
structure. The transitions between the three helical structures are
not visible in $R_{min}$ as the structures are very similar but the
transition to a folded state is strongly marked.

In the following, we examine how the structure of square-well
homopolymers is affected by the monomer size ($\sigma/l$), the range
of the attractive interaction ($\lambda$), and the length of the
polymer chain ($N$).  In particular, we are interested in
understanding the range of parameters where helical structures are
stable.

%%%%%%%%%%%%%%%%%%%%%%%%%%%%%%%%%%%%%%%%%%%%%%%%%%%%%%%%%%%%%%%%%%%%%%%%%%%%%%%%
\subsection{Ratio of monomer size to bond length}

In this section, we study the influence of monomer size, or
equivalently the bond length, on the structure of square-well
homopolymers.  We limit our attention to homopolymers with $N=20$ and
$\lambda=1.5$.  The main effect of changing the monomer size is to
alter the local stiffness of the polymer chain.
Decreasing the size of the monomers (or increasing the bond length)
increases the flexibility of the homopolymer.
The stiffness of a polymer chain can be characterized by the bond
correlation function, which is defined as
\begin{equation}
C(j) = \frac{1}{N-j-1} \sum_{k=1}^{N-j-1}
\frac{\langle\Delta{\bf r}_k\cdot\Delta{\bf r}_{k+j}\rangle}
     {\langle\Delta{\bf r}_k\cdot\Delta{\bf r}_k\rangle}
\end{equation}
where $\Delta{\bf r}_k={\bf r}_{k+1}-{\bf r}_k$ is the orientation of
the $k^{\rm th}$ bond in the polymer, and ${\bf r}_k$ is the position
of the $k^{\rm th}$ monomer.  This function describes the degree to
which the orientations of two bonds are correlated to each other.  The
more flexible the chain, the more rapidly the bond correlation
function decays with the distance $j$ between the bonds.

\begin{figure}[tb]
\begin{center}
\includegraphics[clip,width=\columnwidth]{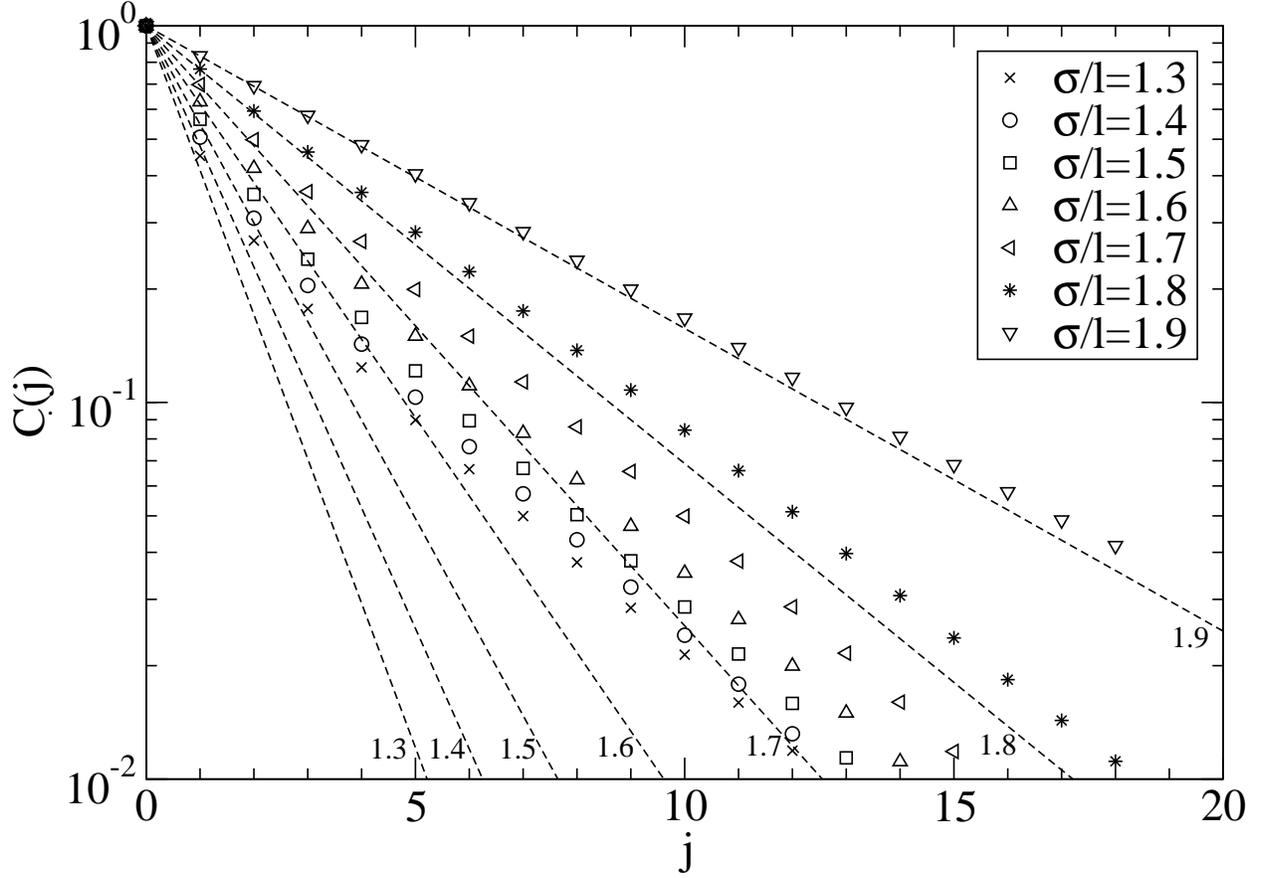}
\caption{\label{fig:bondcorrelation}
Average bond angle correlations for athermal overlapping chains as a
function of separation in a $N=20$ chain.  The dotted lines correspond
to neglecting the influence of long range excluded volume interactions
between the monomers.}
\end{center}
\end{figure}

Figure~\ref{fig:bondcorrelation} presents the bond correlation
functions for athermal chains (i.e.\ $\epsilon=0$) with $N=20$ for
various values of $\sigma/l$.  The symbols are the results obtained
from MD simulations.
The dotted lines are the corresponding exponential decays for the
athermal chains where excluded volume interactions are neglected, with
the exception to those between monomers separated by two bonds, which
give rise to the local stiffness.
At very low values of $\sigma/l$ (not shown), there are no
correlations between the bonds, and the polymer behaves essentially as
a random walk.  
At intermediate values of the overlap parameter, the excluded volume
interactions between monomers separated by several bonds enhance the
correlations between the bonds, and the correlation function decays
algebraically, rather than exponentially.
For $\sigma/l>1.8$, the decay is nearly exponential because the chain
is too stiff for there to be significant excluded volume interactions
between the monomers.

%%%
%%%
%%%
%%%

\begin{figure}[tb]
\begin{center}
\includegraphics[clip,width=\columnwidth]{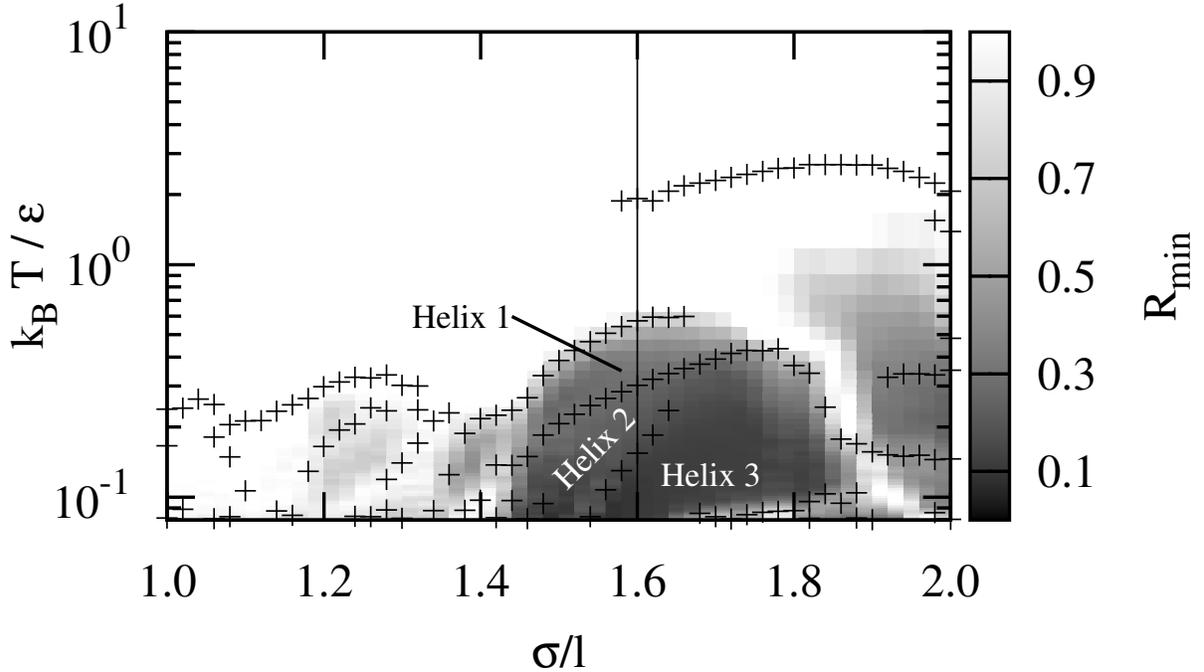}
\caption{\label{fig:overlapsweep}
Diagram of states for isolated square-well homopolymers with $N=20$
and $\lambda=1.5$.  The graph is shaded according to $R_{\rm min}$.}
\end{center}
\end{figure}

A diagram of states for homopolymers with $N=20$ and $\lambda=1.5$ is
given in Figure~\ref{fig:overlapsweep}, which explores the effect of
the monomer overlap parameter ($\sigma/l$).  The crosses mark the
locations of peaks in the heat capacity, and the diagram is shaded
according to the value of $R_{\rm min}$.  The data in
Fig.~\ref{fig:initResults} correspond to the vertical line at
$\sigma/l=1.6$ in Fig.~\ref{fig:overlapsweep}.

There are what at first appear to be discontinuities in the heat
capacity maxima in the diagram.  These peaks are generally weak maxima
in the heat capacity which are hidden behind the rapid increase in
$C_v$ due to the presence of sharper peaks at another temperature.
The highest temperature maximum in $C_v$ typically corresponds to the
coil-globule collapse or ``theta point'' (see point $d$ in
Fig.~\ref{fig:initResults} ).

%%% 
%%% 
%%% 
%%% 

Homopolymers with $\sigma/l\lesssim1.1$ crystallize into compact,
nearly spherical, regularly packed structures at low temperatures
\cite{Zhou_etal_1996,Zhou_etal_1997}.  The comparatively high value of
the RMSD for these polymers, however, indicates that the structures
that they freeze into are not unique.  There may be several
arrangements of the bonds of the polymer for a given ``crystalline''
packing of the monomers.  Consequently, these polymers are like
glasses at low temperatures.

When $\sigma/l\gtrsim1+\delta/l=1.1$, directly bonded monomers always
overlap one another.  At low temperatures, homopolymers with
$1.1\lesssim\sigma/l\lesssim 1.4$ (see Fig.~\ref{fig:overlapsweep})
exhibit a ``freezing'' transition, but, similarly to the polymers with
$\sigma/l\lesssim1.1$, they do not lock into a single, stable
conformation.  The high value of the RMSD indicates that many folded
configurations exist. On visual inspection of these configurations,
helical features are visible within some other structure.  For
example, the ends of the polymer may be wrapped around the outside of
a helical core. These ``loose ends'' increase the number of possible
frozen states and therefore increase the value of $R_{min}$.

For homopolymers with a well width of $\lambda=1.5$, monomers
separated by two bonds are permanently within each others attractive
wells when $\sigma/l>2(1+\delta/l)/\lambda\approx1.47$.  This
coincides with the onset of the region of low values for $R_{\rm
  min}$, where homopolymers fold into the helix 1, 2, and 3
structures.
%are defined by the characteristic RMSD of the folded configuration.
Here, the homopolymers fold into a single, helical conformation
(neglecting the distinction between the left and right handed
configurations).
A significant portion of the folded parameter space is occupied by the
helix 3 structure, which is the most rigid of the helix structures.

At high overlaps, the values of $R_{\rm min}$ are on average lower due
to the stiffness of the chain limiting the range of motion of the
monomers.  There is a sharp transition at $\sigma/l\approx 1.8$ with
an increase in $R_{\rm min}$ along the line of the heat capacity
peaks.  For polymers with a well width of $\lambda=1.5$, two monomers
separated by 4 bonds cannot interact with each other when
$\sigma/l>\sqrt{7/2}\approx 1.87$ \cite{Magee_etal_2008}.
If we account for the fact that in the simulations the bonds can
stretch by $10\%$, then this would occur at $\sigma/l\gtrsim1.70$,
which coincides with loss of the helix 1 structure.

It appears that the observed helix structures are closely related to
the constraint of interactions between monomers in the chain.  The
values at which certain interactions become prohibited depends on the
well width $\lambda$, and the effect of this parameter is explored in
the next section.

%%%%%%%%%%%%%%%%%%%%%%%%%%%%%%%%%%%%%%%%%%%%%%%%%%%%%%%%%%%%%%%%%%%%%%%%%%%%%%%%
\subsection{Range of attractive interaction}

\begin{figure}[tb]
\begin{center}
\includegraphics[clip,width=\columnwidth]{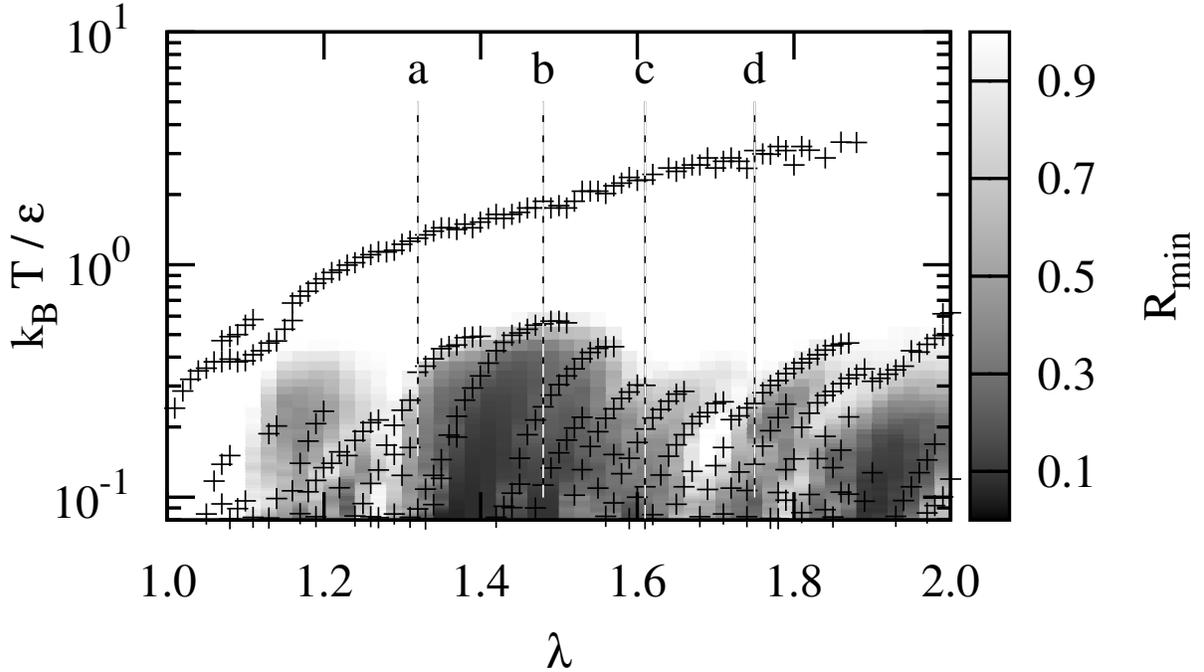}
\caption{\label{fig:lamsweep}
Diagram of states for isolated square-well homopolymers with $N=20$
and $\sigma/l=1.6$.  The graph is shaded according to $R_{\rm min}$.
The letters and dashed lines correspond to the configurations shown in
Fig.~\ref{fig:lamsweepsnaps}. }
\end{center}
\end{figure}

\begin{figure}[tb]
\begin{center}
\includegraphics[clip,width=\columnwidth]{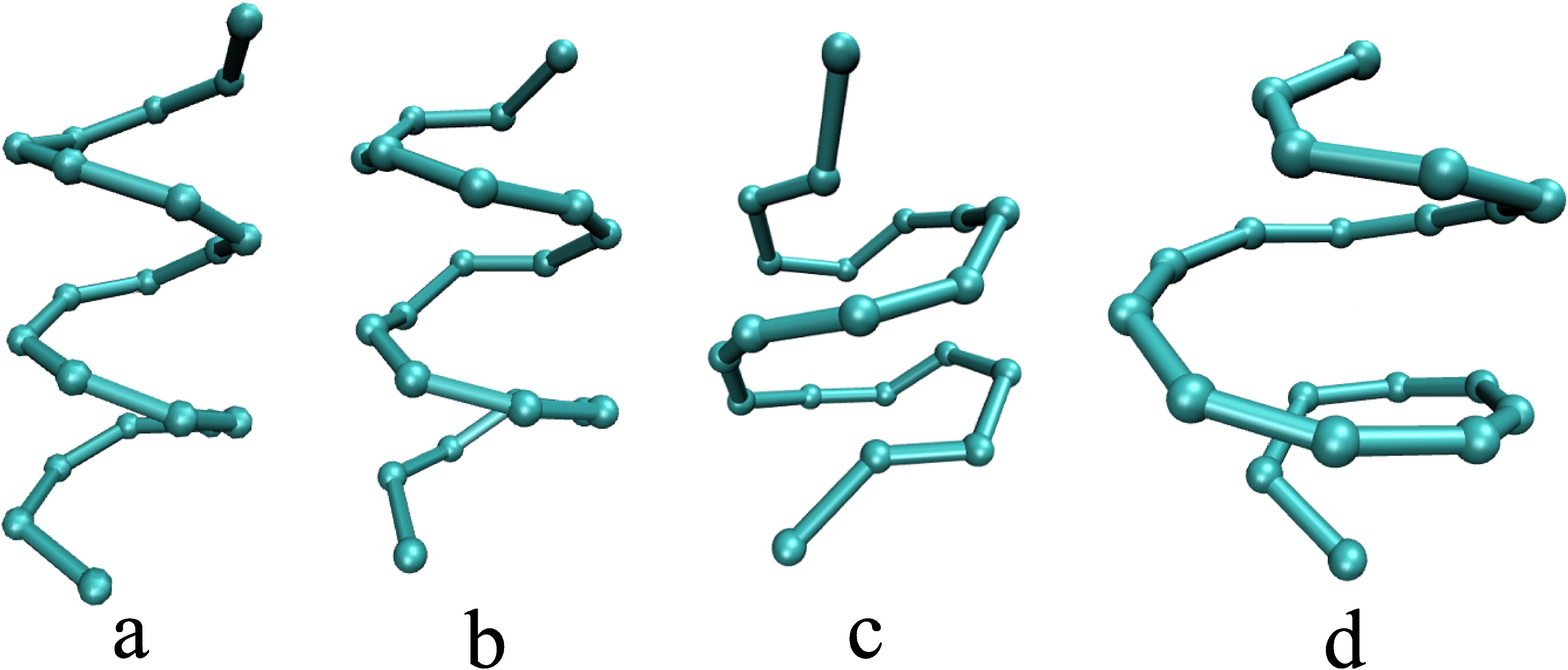}
\caption{\label{fig:lamsweepsnaps}
Representative configurations at the state points indicated in
Fig.~\ref{fig:lamsweep}.}
\end{center}
\end{figure}

Now, we examine the influence of the range of the attractive
interaction, which is characterized by the parameter $\lambda$.  In
this section, we limit the analysis to square-well homopolymers with
$N=20$ and $\sigma/l=1.6$.  A diagram of states is provided in
Fig.~\ref{fig:lamsweep}.  Several sample configurations are presented
in Fig.~\ref{fig:lamsweepsnaps} at various values of $\lambda$.
From the diagram and the associated configurations, we see that a
series of distinct, helical structures are formed at low temperatures.
The range of the attractive interaction appears to control the radius
of the helix: smaller well widths lead to helical structures with a
smaller radius and a larger pitch.

%%% 
%%% 
%%% 
%%%

%%% 
%%% 
%%% 
%%%

At low values of the well width ($\lambda\lesssim1.3$), helical
structures appear with structural variations, much like what occurs at
low values of the overlap parameter $\sigma/l$.
For high values of $\lambda$, the helix structure begins to degrade.
It still retains the spiral structure, however, it no longer has a
constant radius.  Interestingly, for the structure shown in
Fig.~\ref{fig:lamsweepsnaps}d, the monomers appear to be packed in a
fairly spherical crystalline arrangement.
If the well width becomes too large, then the helix structure will
completely vanish, replaced by another structure.

For an overlap of $\sigma/l=1.6$ and a well width of
$\lambda\ge1.375$, monomers separated by two bonds are permanently
within each others attractive well.  This again coincides with a large
decrease in $R_{\rm min}$, indicating a single stable structure.  As
with the diagram of states in the overlap parameter $\sigma/l$ (see
Fig.~\ref{fig:overlapsweep}), it is easy to distinguish certain
helical structures using $R_{\rm min}$.  

It is interesting to note that the helices observed here all have a
much higher monomer per turn count than the alpha helix commonly found
in nature. There are 4 residues per turn of the alpha helix, whereas
the wider helices presented here contain 7 for the tightest helix
observed (Fig.~\ref{fig:lamsweep}a). Maritan et al.\
\cite{Maritan_etal_2000} characterized their compact string helices
using a parameter $f$, related to the helix radius and monomer spacing
in consecutive turns of the helix. Applying their analysis, the values
of $f$ exhibited by our helices are consistently above the value of
$f\approx1$ (e.g., Fig.~\ref{fig:lamsweep}a--d $f\approx1.2$, $1.1$,
$1.3$, and $1.1$, respectively) reported by Maritan for compact
strings and naturally occurring alpha helices. These larger values of
$f$ may be due to the manner in which we introduce stiffness (i.e.,
overlapping spheres).

Unlike the overlap parameter, the transitions between the various
helical states are typically marked by peaks and large changes in
$R_{\rm min}$, as the well width parameter $\lambda$ has a significant
effect on the structure of the folded state.  In the following
section, we explore the structure as a function of chain length.

%%%%%%%%%%%%%%%%%%%%%%%%%%%%%%%%%%%%%%%%%%%%%%%%%%%%%%%%%%%%%%%%%%%%%%%%%%%%%%%%
\subsection{Chain Length
\label{sec:length}}

For the square-well homopolymer, the main driving force for the
formation of the helix is the tendency of the polymer to recover
interaction energy through the contacts of its constituent monomers.
This energetic driving force is balanced against the loss of entropy
encountered in restricting the polymer to the helical structure (to
maintain the necessary contacts).  If the polymer chain is too short,
then the energy recovered will not be sufficient to overcome the
entropy loss, and the helix will not be stable.  If the polymer chain
is too long, then structures other than the helix are expected to be
stable.  Therefore, we expect the helix to appear only within a window
of chain lengths.  In this section, we examine the range of $N$ where
the helical structure is stable.

\begin{figure}[tb]
\begin{center}
\includegraphics[clip,width=\columnwidth]{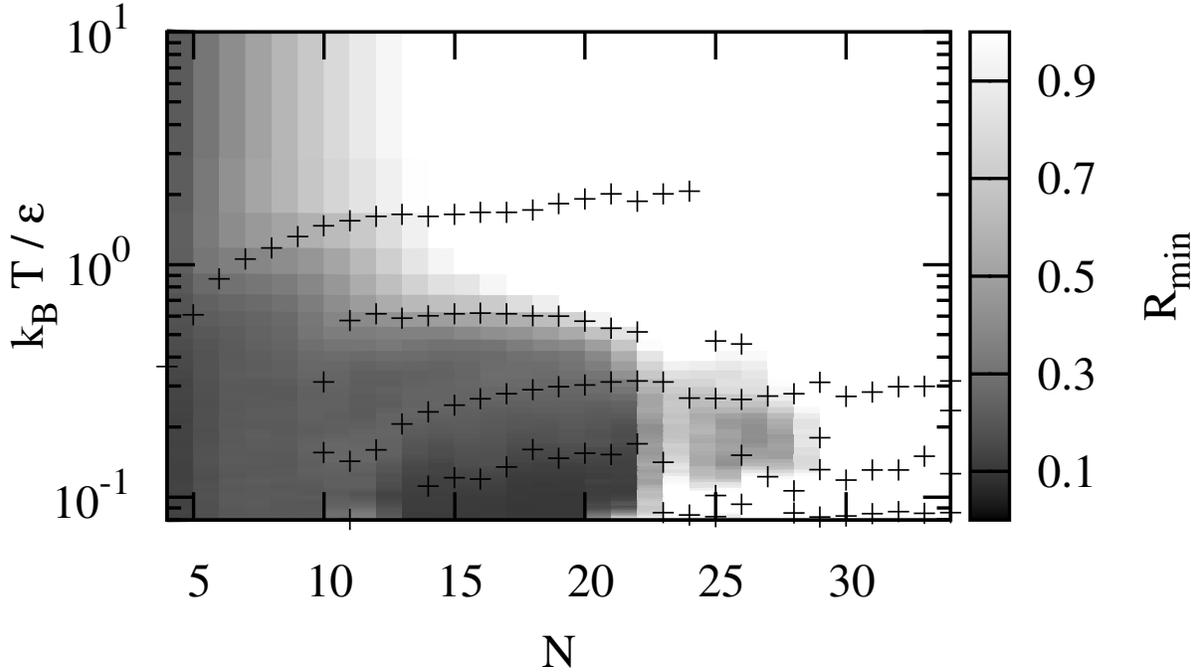}
\caption{\label{fig:sig16lam15}
Diagram of states for isolated square-well homopolymers with
$\sigma/l=1.6$ and $\lambda=1.5$ as a function of the chain length
$N$.  The graph is shaded according to $R_{\rm min}$.}
\end{center}
\end{figure}

The diagram of states for square-well polymers with $\lambda=1.5$ and
$\sigma/l=1.6$ is presented in Fig.~\ref{fig:sig16lam15}. For small
chain lengths ($N \lesssim 12$) the RMSD is, on average, a low
value. This is due to the short distance that monomers can actually be
separated in space.
This can be accounted for by reducing $R_{min}$ by the chain length;
however, similar structures at different chain lengths typically
exhibit the same value of $R_{min}$ and this data would be lost.
The conditions where helical structures are formed are still well
defined by the heat capacity peaks and areas where the value of
$R_{\rm min}$ is low.  For this system, the chain must consist of at
least $N=10$ monomers before helices can form.  The helix 3 structure
does not appear until $N=14$, and the largest chain length at which
the helix structure is stable is $N=22$.

%%%
%%%
%%%
%%%

\begin{figure}[tb]
  \begin{center}
    \includegraphics[clip,width=0.9\columnwidth]{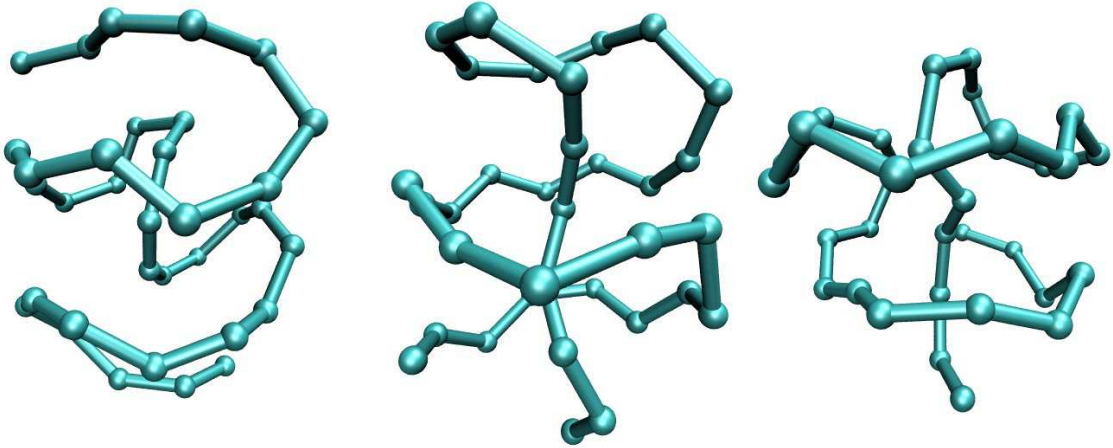}
    \caption{\label{fig:sig16lam15snap}
Samples of stable configurations for square-well homopolymers with
$\sigma/l=1.6$, $\lambda=1.5$, and $N=34$ at $k_BT/\epsilon=0.13741$.}
  \end{center}
\end{figure}

At low temperatures, homopolymers with $N>22$ appear to freeze into
rigid structures, yet the high values of $R_{\rm min}$ indicate that
the homopolymer does not freeze into a single, repeatable, folded
structure.  In fact, these folded states are no longer unique, and
several distinct structures exist with comparable free energies. These
longer homopolymers arrange themselves into regularly packed
structures with a spherical shape.
Figure~\ref{fig:sig16lam15snap} provides several snapshots of
configurations for square-well homopolymers with $N=34$ at a
temperature $k_BT/\epsilon=1.35$ (see also Fig.~\ref{fig:sig16lam15}).
These chiral structures have the same interaction energy, and they are
all stable over long times.  They appear to be variations on a similar
structural theme: a core of a few monomers with a chiral outer core.

%%%
%%%
%%%
%%%

The RMSD's of the different folded structures in
Fig.~\ref{fig:sig16lam15snap} lie between $0.86\sigma$ and
$1.05\sigma$, which is a relatively high value.  Thus, the RMSD can
discriminate between distinct folded structures, provided that the
configurations within each of the structures have a low average RMSD.
If we perform a quality threshold (QT) cluster analysis
\cite{Heyer_etal_1999} of the RMSD between all pairs of sample
configurations using a cutoff value of $<0.25\sigma$ to group the data
and a threshold of 1\% to eliminate intermediates, we can attempt to
count the number of distinct structures formed.  We perform this
counting at the heat capacity minima, as the heat capacity maxima tend
to occur at transitions between structures.  For $N=22$, only one
cluster is apparent, which indicates that the homopolymer folds into a
unique structure at low temperatures; in this case, it is a helix.
In contrast, for $N=23$, a single helix occupies approximately 50\% of
the simulation snapshots.  The remainder are a large number of
variations on the helix with ``loose'' ends wrapped around the central
coil.

In fact, once the single helix structure is no longer dominant the
number of distinct folded structures rapidly increases with the length
of the homopolymer.  These polymers will behave similarly to a glass
at low temperatures, becoming trapped into one of these many
structures.

\begin{figure}[t]
  \begin{center}
    \includegraphics[clip,width=\columnwidth]{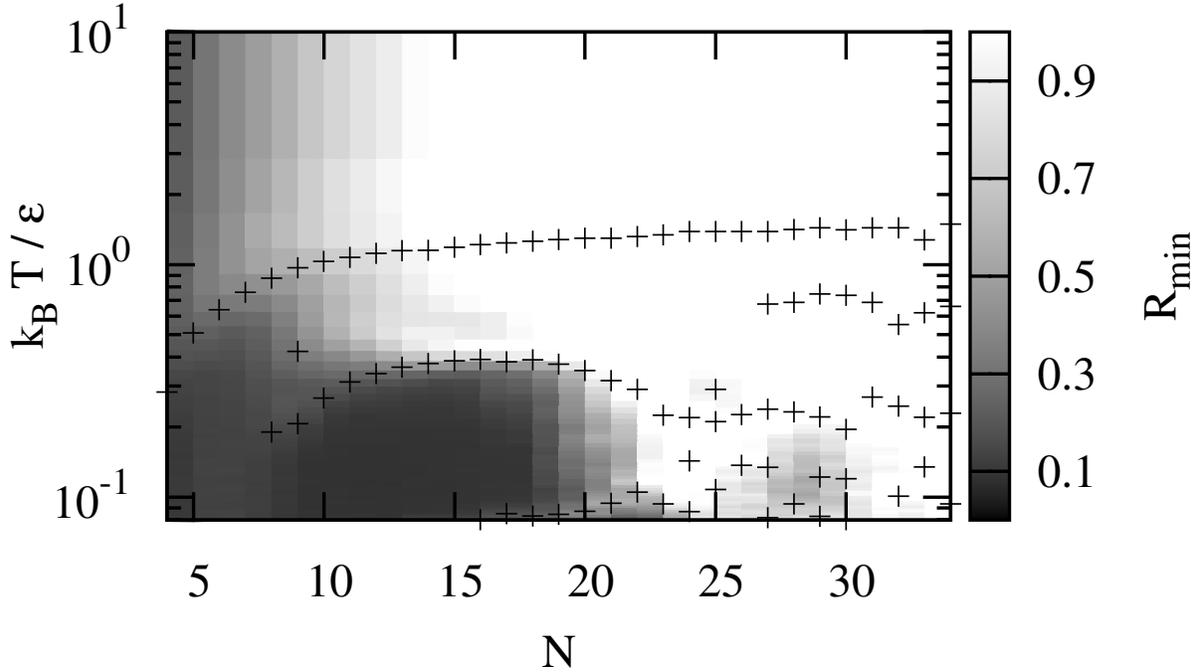}
    \caption{\label{fig:sig16lam132}
Diagram of states for isolated square-well homopolymers with
$\sigma/l=1.6$ and $\lambda=1.32$ as a function of the chain length
$N$.   The graph is shaded according to $R_{\rm min}$.}
  \end{center}
\end{figure}

To understand how the range of the attractive interactions affects the
window of chain lengths where the helix is stable, we examine
square-well homopolymer chains with $\lambda=1.32$ and $\sigma/l=1.6$.
The diagram of states is presented in Fig.~\ref{fig:sig16lam132}.
These polymers tend to form helices at shorter chain lengths than
polymers with a wider well widths (cf.\ Fig.~\ref{fig:sig16lam15} for
$\lambda=1.5$).
The helices formed by the $\lambda=1.32$ polymers have a tighter
radius and are more rigid (lower value of $R_{\rm min}$) than the
helices formed by the $\lambda=1.5$ polymers.  The shortest
homopolymer that forms a helix ($N=8$) appears to be correlated to the
number of monomers in a single turn of the helix.  The helix structure
vanishes for chain lengths greater than $N=22$, which is similar to
what is found for homopolymers with $\lambda=1.5$.  At longer chain
lengths, the system again exhibits multiple folded states, and the
structures formed are similar to those displayed in
Fig.~\ref{fig:sig16lam15snap}.

%%%
%%%
%%%
%%%

\begin{figure}[tb]
\begin{center}
\includegraphics[clip,width=\columnwidth]{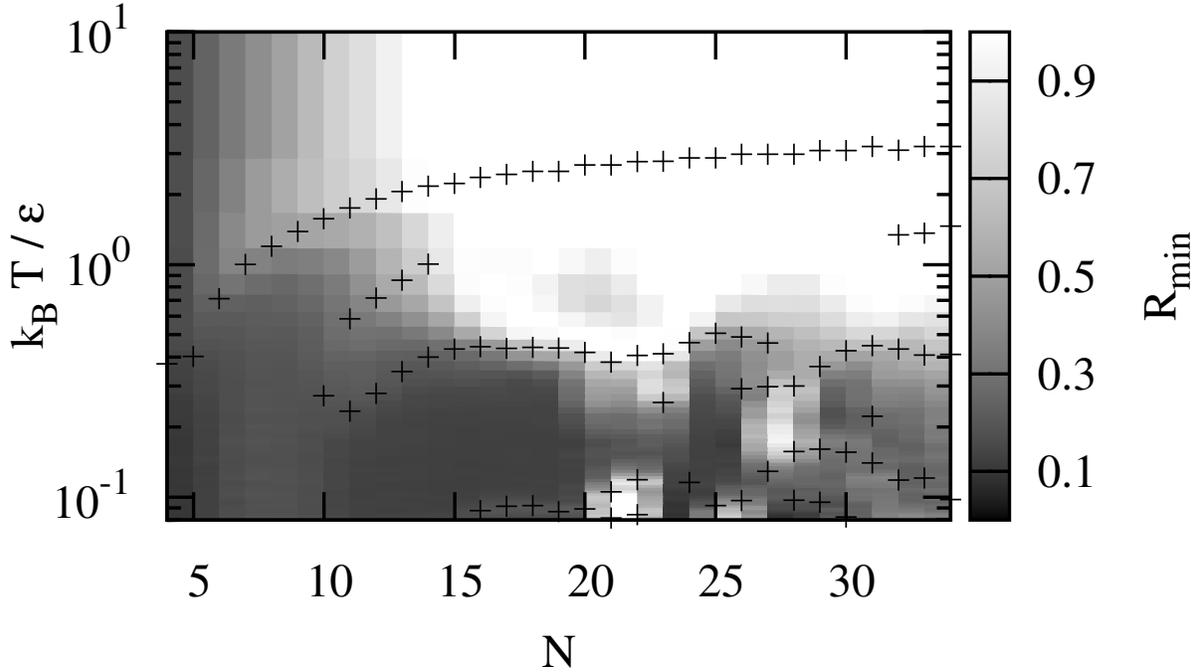}
\caption{\label{fig:sig18lam15}
Diagram of states for isolated square-well homopolymers with
$\sigma/l=1.8$ and $\lambda=1.5$ as a function of the chain length
$N$.   The graph is shaded according to $R_{\rm min}$.}
\end{center}
\end{figure}

To investigate the influence of the monomer size (or bond length) on
the window of chain lengths where the helix is formed, we examine
square-well homopolymers with $\sigma/l=1.8$ and $\lambda=1.5$.  The
diagram of states for these systems is given in
Fig.~\ref{fig:sig18lam15}, which shows a rich range of structural
behavior.  The minimum chain length for helix formation is larger
($N=11$) than for polymers with a monomer size of $\sigma/l=1.6$. The
increased stiffness of the chain is limiting the curvature of the
helix formed, thus requiring more monomers per turn of the helix.  It
appears that the typical ``glassy'' behavior of the longer polymers
has been eliminated for the examined chain lengths.  Therefore, the
maximum chain length displaying a helical structure must be determined
using QT analysis and visual inspection.  The last chain length where
a single helix structure is stable is $N=22$, yet for the longer chain
lengths ($23\le N\le34$), the folded structures remain unique and not
glassy.  A single structure, which we refer to as the ``barbers pole''
structure, is observed over these chain lengths and is similar to the
two rightmost structures of Fig.~\ref{fig:sig16lam15snap}.  This
structure was first observed by Magee et al.\ (See Fig.~2 of
Ref.~\cite{Magee_etal_2006}).  Unlike the configurations of
Fig.~\ref{fig:sig16lam15snap}, the polymer is too stiff to allow the
reversal of direction or doubling back of the outer spiral in the
``barbers pole''.  It appears that the increased stiffness has reduced
the number of possible low energy permutations, which manifest in more
flexible chains as the doubling back of the outer spiral, to a single
configuration.  The small regions of high $R_{min}$ at low
temperatures in Fig.~\ref{fig:sig18lam15} correspond to broad peaks in
the heat capacity where transitions between different ``barbers pole''
structures occur.

Chains with a higher value of $\sigma/l$ appear to favor a single
folded structure at longer chain lengths than compared to more
flexible chains.  This is understandable as in the limit of a rigid
chain there is only one possible physical configuration.  As the chain
becomes stiffer the number of low energy permutations on a structural
theme are limited until only one configuration becomes optimal.

%%%
%%%
%%%
%%%

%%%%%%%%%%%%%%%%%%%%%%%%%%%%%%%%%%%%%%%%%%%%%%%%%%%%%%%%%%%%%%%%%%%%%%%%%%%%%%%%
%%%%%%%%%%%%%%%%%%%%%%%%%%%%%%%%%%%%%%%%%%%%%%%%%%%%%%%%%%%%%%%%%%%%%%%%%%%%%%%%
\section{Conclusions
\label{sec:conclusions}} 

In this work, we have used event-driven molecular dynamics, coupled
with the replica exchange and histogram reweighting techniques, to
explore the behavior of isolated, square-well homopolymers.
The structural properties of these polymers were characterized through
a combination of configurational snapshots, monomer contact maps, and
the root mean square deviation of the configuration, combined with QT
cluster analysis.  The RMSD is able distinguishing between the
unfolded and folded helix states.  QT cluster analysis of the RMSD
allows the estimate of the number of folded states, which reflects the
``variability'' of the state.

The homopolymer model studied here exhibits complex behavior.  The
stability of the helix structure is related to the
constriction/elimination of interactions between monomers separated by
a number of bonds in the chain.  This is affected by the chain
stiffness, which controlled by the monomer overlap parameter
$\sigma/l$.
The pitch and curvature of a helix is governed mainly by the range of
the attractive interaction $\lambda$.  Helices form with a higher
curvature for short-range attractive wells.  For larger values of
$\lambda$, the monomers pack into a more spherical arrangement while
still retaining a spiral bond structure.

Helices are only stable within a window of the chain length $N$.  The
lower limit appears to be related to the number of monomers in a
single turn of the stable helix structure.  Above a critical chain
length, the isolated homopolymer folds into a rapidly increasing
number of stable states, displaying characteristics reminiscent of a
glass transition.  These structures are more compact and spherical
than their lower $N$ counterparts, result from a minimization of the
surface area to volume ratio of the polymer.

Finally, as the stiffness of the homopolymer is increased ($\sigma/l
\to 2(1+\delta/l)$) the number of observed folded states in longer
chain lengths is reduced.  At an overlap of $\sigma/l=1.9$, we only
observe unique folded states for the range of polymer lengths studied
($4\le N\le34$).

%%%%%%%%%%%%%%%%%%%%%%%%%%%%%%%%%%%%%%%%%%%%%%%%%%%%%%%%%%%%%%%%%%%%%%%%%%%%%%%%
%%%%%%%%%%%%%%%%%%%%%%%%%%%%%%%%%%%%%%%%%%%%%%%%%%%%%%%%%%%%%%%%%%%%%%%%%%%%%%%%
\begin{acknowledgments}
  MN Bannerman acknowledges support from an EPSRC DTA. 
\end{acknowledgments}

%%%%%%%%%%%%%%%%%%%%%%%%%%%%%%%%%%%%%%%%%%%%%%%%%%%%%%%%%%%%%%%%%%%%%%%%%%%%%%%%
%%%%%%%%%%%%%%%%%%%%%%%%%%%%%%%%%%%%%%%%%%%%%%%%%%%%%%%%%%%%%%%%%%%%%%%%%%%%%%%%

%%%%%%%%%%%%%%%%%%%%%%%%%%%%%%%%%%%%%%%%%%%%%%%%%%%%%%%%%%%%%%%%%%%%%%%%%%%%%%%%
%%%%%%%%%%%%%%%%%%%%%%%%%%%%%%%%%%%%%%%%%%%%%%%%%%%%%%%%%%%%%%%%%%%%%%%%%%%%%%%%
%%%%%%%%%%%%%%%%%%%%%%%%%%%%%%%%%%%%%%%%%%%%%%%%%%%%%%%%%%%%%%%%%%%%%%%%%%%%%%%%
\end{document}